\documentclass[pra,showpacs,amssymb,twocolumn,footinbib]{revtex4}

\usepackage{amsmath}
\usepackage{amsfonts}
\usepackage{graphicx}

\def\>{\rangle}
\def\<{\langle}
\def\be{\begin{equation}}
\def\ee{\end{equation}}

\begin{document}
\title{On the communication complexity of establishing a shared
reference frame}

\author{Terry Rudolph}
\author{Lov Grover}
\affiliation{Bell Labs, 600 Mountain Ave., Murray Hill, NJ 07974,
U.S.A.}

\date{\today}

\begin{abstract}

We discuss the communication complexity of establishing a shared
reference frame, in particular examining the case of aligning
spatial axes via the exchange of spin-1/2 particles. Unlike
previous work we allow for multiple rounds of communication, and
we give several simple examples demonstrating that nontrivial
tradeoffs between the number of rounds and the type of
communication required exist. We then give an explicit protocol
for aligning spatial axes via the exchange of spin-1/2 particles
which makes no use of either exchanged entangled states or of
joint measurements. Rather it works by performing a simple type of
distributed quantum computation. To facilitate comparison with
previous work, we show that this protocol achieves a worst case
fidelity for the much studied problem of ``direction finding''
that is asymptotically equivalent (up to polylog factors) to the
optimal average case fidelity achievable via a single forward
communication of entangled states.

\end{abstract}


\maketitle

Quantum physics allows for powerful new communication tasks that
are not possible classically, such as secure secret key
distribution~\cite{BB84} and entanglement-enhanced classical
communication~\cite{Mac02}. Such quantum communication tasks
generically require one party to prepare systems in well defined
quantum states, and to send these systems to another party. Since
the states used are generally defined only with respect to some
sort of reference frame, a perfect shared reference frame (SRF)
between both parties is normally presumed. In general, however,
establishing a perfect SRF requires infinite communication (i.e.
transmitting a system with an infinite-dimensional Hilbert space,
or an infinite number of systems with finite-dimensional Hilbert
spaces).  In practice, perfect SRFs are an idealization, and any
finite (i.e., approximate) SRF should be viewed as a quantitative
physical resource, since, along with requiring communication to
establish, quantum mechanics dictates that finite SRFs necessarily
drift~\cite{Wig57} and thus are intrinsically depleted over time.
Moreover, any finite SRF that is treated quantum mechanically will
inevitably suffer disturbances during measurements, again
depleting the SRF. We also note that shared prior entanglement, a
valuable resource in quantum information theory, can be consumed
to establish SRFs \cite{tez1}.


 In quantum communication theory, the specific physical systems being
exchanged determine the type of reference frame that the
communicating parties must share; conversely, the ability to
exchange physical systems generally allows for certain reference
frames to be established. For example, in order for two parties to
agree on the superposition $\alpha|\uparrow\>+\beta|\downarrow\>$
of a single spin-1/2 system, they must share aligned spatial axes;
conversely, by exchanging spin-1/2 systems they can establish
aligned spatial axes.

The problem of using spin-1/2 systems to establish either a single
direction in space or an orthogonal trihedron (xyz-axes), has
received considerable attention
\cite{Mas95,Lat98,Gis99,massar,Per01b,Bag00}. In particular, the
following \emph{standard scenario} has been studied in depth:
Alice sends Bob $N$ spin-1/2 particles in a state which encodes
some spatial direction $\vec{n}_A$. Bob performs a measurement on
the $N$ spins, which results, with probability
$P(\vec{n}_e|\vec{n}_A)$, in an estimation $\vec{n}_e$ as to the
direction $\vec{n}_A$. The fidelity of the estimation is defined
as $\tfrac{1}{2}(1+\vec{n}_e\cdot\vec{n}_A)$, and the goal is to
optimize the \emph{average} fidelity. That is, for uniformly
chosen $\vec{n}_A$ one tries to maximize the expected fidelity,
$\bar{F}$, with respect to initial states prepared by Alice and
measurements performed by Bob. (Note that a random guess of
direction has an expected fidelity of 1/2; a fidelity of 0
corresponds to an estimate antiparallel to $\vec{n}_A$). In
general it is found that if Alice sends Bob the systems in a
tensor product of pure states, then $\bar{F}^{\max}=
1-O(\frac{1}{N})$, while if Alice prepares entangled states then
$\bar{F}^{\max}= 1-O(\frac{1}{N^2})$. In both cases the
measurements Bob must perform to achieve this are \emph{joint}
(i.e. entangled) measurements over the $N$ particles, and in
general they are positive operator valued measurements (POVM's) as
opposed to standard von-Neumann projection valued measurements
(PVM's). It is sometimes claimed \cite{Bag00} that more general
encoding of a spatial direction in entangled states can achieve
$\bar{F}^{\max}= 1-O(\frac{1}{2^N})$, however such encodings can
only be performed if Alice and Bob \emph{already} share aligned
spatial axes (as was noted in \cite{Bag00}), and in this case
Alice can do no better than to use the $N$ particles to send Bob
\emph{classical} bits of information specifying an approximation
to $\vec{n}_A$.\footnote{It appears that in quant-ph/0303019 this
mistake has been made again. For the procedures outlined in that
paper, it seems to us that Alice and Bob need to share aligned
spatial axes - in which case they can do no better than to simply
use standard dense coding of classical information.} We will
always assume here that Alice and Bob do not, a-priori, share any
sort of spatial reference frame.

There are several ways in which this standard scenario (and the
extension of it in which Alice and Bob align an orthogonal
trihedron as opposed to a single direction) is somewhat
unsatisfactory. For a start, the particular choice of cost
function (e.g. the fidelity) has a strong bearing on what the
optimal states and measurements turn out to be \cite{massar}.
Secondly, the optimizations are performed for the \emph{average
case} scenario, and not the \emph{worst case} scenario, which is
arguably more interesting (and which is the norm for communication
complexity type problems). This yields the difficulty that if we
wish to ask questions pertinent to future quantum communication
using spatial axes aligned under such a procedure, it is somewhat
problematic to translate these results into standard properties of
the quantum channel. This in turn makes it difficult to determine
the extent to which such communication overhead can be amortized.
The standard scenario also ignores the question as to whether
allowing backwards communication (from Bob to Alice) can improve
their ability to align their reference frames. Finally, in quantum
communication scenarios it is natural to presume that Alice and
Bob have access to \emph{both} classical and quantum channels, and
to examine the extent to which classical and quantum communication
can in some sense be traded off against each other.\footnote{We
presume that classical bits are communicated in such a manner that
they convey no information suitable for aligning spatial reference
frames!} Peres has raised some interesting questions about
classical communication costs within the standard scenario
\cite{peresprivate}, although within this scenario, from a
communication theory perspective, one may generally assume that
such costs are amortized into the definition of the protocol.
Below we will give some simple examples of protocols for which
such amortization not possible.

In this paper our aim is expand the analysis of procedures for
establishing SRFs, by demonstrating the wealth of nontrivial
possibilities which remain to be explored. We also approach the
problem with a view to rectifying some of the shortcomings of the
standard scenario mentioned above. As such, we  consider
strategies for aligning a spatial reference frames that allow Bob,
within a worst case scenario, to directly determine the Euler
angles which relate his and Alice's reference frames. More
precisely, if $\theta$ is an Euler angle relating Alice and Bobs'
axes, and $\theta'$ is the estimation of $\theta$ inferred by Bob,
then we will be interested in the amount (and type) of
communication required for protocols that achieve
$Pr[|\theta-\theta'|\ge \delta]\le \epsilon$, for some fixed
$\epsilon,\delta>0$. By setting $\delta=1/2^{k+1}$ we say that
with probability $(1-\epsilon)$ Bob has a $k$ bit approximation to
$\theta$. In order to connect with known results for the standard
scenario, we will show that the particular procedure we propose
for determining the Euler angles can be used to give a \emph{worst
case} fidelity of $F = 1-O(\frac{\log^2 N}{N^2})$, which is within
a logarithmic factor of the best average case fidelity obtainable
in the standard scenario. However, in contrast to the standard
scenario procedure which achieves this best average case fidelity,
the protocol that we propose makes no use of entanglement - either
in the states that must be prepared or in the measurements that
must be performed. We feel this is of great pragmatic importance,
since if Alice and Bob had the ability to create and exchange the
arbitrarily large entangled states, and perform the arbitrarily
large joint measurements, required within the standard scenario,
then in most situations they would be far better off to use the
ideas presented in \cite{BRS} - wherein it is shown how they can
perform quantum communication \emph{perfectly} (i.e without having
noise due to the finiteness of the SRF) and with asymptotically no
loss of resources.

Before presenting our specific protocol, we discuss a few simple
examples, designed to indicate the diversity of options that open
up once we consider bi-directional communication of both classical
and quantum bits, and to show that we should expect, in general,
some highly non-trivial tradeoffs  - as well as classical
communication that cannot be amortized. For simplicity let us
assume that Bob is trying to estimate the direction of Alice's $z$
axis. With a single qubit of forward communication (the standard
scenario), Alice sends Bob a single spin-1/2 in the state
$|z_A^+\>$ (for almost all of this paper we assume the qubits are
spin-1/2 particles). Although we always assume it is Bob who must
estimate the direction, the same fidelity can be achieved by one
qubit of \emph{backward} communication from Bob to Alice, followed
by one forward bit of \emph{classical} information from Alice to
Bob. This is done by Bob preparing two spins in a singlet state,
and sending one of the spins to Alice. Alice performs a
measurement on the spin, which steers \cite{erwin} its partner
being held by Bob to either $|z_A^+\>$ or $|z_A^-\>$ - she then
sends a classical bit to inform Bob of the outcome.

If we consider two qubits worth of communication, it is known that
Alice would prefer Bob to be end up with an \emph{antiparallel}
pair of spins \cite{Gis99}. This can be achieved with one qubit of
backward communication, followed by one qubit and one classical
bit of forward communication. To do this we simply modify the
procedure mentioned above, so that after Alice has made her
measurement, in addition to the classical bit she also sends an
extra qubit aligned anti-parallel to her measurement outcome. Note
that with two qubits of backward communication, implementing a
similar procedure would result in Bob's 2 qubits being in one of
the 4 pairs of states $|z_A^+z_A^+\>, |z_A^-z_A^-\>,
|z_A^+z_A^-\>, |z_A^-z_A^+\>,$ with equal likelihood. Alice would
need to send two classical bits to Bob, and moreover cannot ensure
that the qubits are antiparallel.

With two qubits of communication there is yet another option
available to Alice and Bob. Instead of measuring her half of the
singlet, Alice can simply apply some unitary operation to it and
return it to Bob. Because of the differences in their reference
frames, the entangled state held by Bob will now encode some
information about their relative axes alignment. A detailed
analysis of such a procedure can be found in \cite{vidal}. It is
interesting to note that if Alice does a $\sigma^z_A$ rotation,
and returns the entangled qubit along with another spin which is
aligned with her $z$-axis, then this procedure (which has involved
one backward and two forward qubits of communication) gives an
average fidelity the same as if Alice had sent three parallel
spins to Bob. As such it is not, a-priori, particularly
interesting. However a significant difference arises in the
measurement Bob must perform on the three spins he now holds. It
is easy to show that he can achieve this fidelity by performing a
Bell measurement on the two entangled spins (one of the Bell
outcomes has probability 0 of occuring) and a standard PVM on the
remaining spin - the specific nature of which is based on the
outcome of the Bell measurement. Thus, the same average fidelity
is achieved by a PVM with classical feedforward - a decidedly
different and simpler measurement than the minimal and optimal
POVM known for the standard scenario \cite{Lat98}.

We now turn to the simplest protocol we have been able to find for
determining the Euler angles $\{\phi,\theta,\psi\}$ in a
worst-case scenario. Unless otherwise indicated by a
subscript/superscript, all states and operators are written in
Bob's frame of reference. We define the Euler angles such that the
rotation matrix describing the change from Alice to Bobs' frame of
reference is given by $R\equiv
e^{-i\psi\sigma_z/2}e^{-i\theta\sigma_y/2}e^{-i\phi\sigma_z/2}$.
Explicitly,
\[
R=e^{-i(\psi +\phi )/2}\left(
\begin{array}{cc}
\cos \theta /2 & -e^{i\phi }\sin \theta /2 \\
-e^{i\psi }\sin \theta /2 & e^{i(\phi +\psi )}\cos \theta /2%
\end{array}%
\right) .
\]

Let $\theta=\pi T$, where $0\le T\le 1$ has a binary expansion
$T=0\cdot t_1t_2\ldots$. The protocol we propose involves Alice
and Bob following an iterative procedure which determines the bits
$t_1$, $t_2$ up to $t_k$ independently. We choose the probability
of error for each bit of $T$ to be $ \epsilon /k,$ so that after
finding the first $k$ bits of $T$ the total probability of error
is $1-(1-\epsilon /k)^{k}$ $\leq \epsilon. $

To find $t_1$, Alice sends a single spin polarized in her $z$
direction, and Bob measures it in his $\pm z$ basis. The
measurement by Bob yields the outcome 1 (spin down say) with
probability: \[ P_{1}=\cos^2\tfrac{\theta}{2}=\tfrac{1}{2}[1+\cos
2\pi T ]=\tfrac{1}{2}[1+\cos (2\pi\;0\cdot t_1t_2\ldots)]. \]
Repeating this $n$ times, Bob obtains an estimate $P_1'$ as to the
true value of $P_1$, and thus an estimate $T'$ of the true value
of $T$. If we choose $n$ (details below) such that $|P_1-P_1'|\le
1/4$ with probability $(1-\epsilon/k)$, then $|T-T'|\le 1/4$ with
the same probability, and this implies that $T'$ agrees with $T$
to at least the first bit $t_1$.

We now show how the above process for estimating the first bit of
$T$ can be generalized so as to estimate the $(j+1)$'th bit of
$T$. Consider the situation where Bob sends a qubit in the state
$|z^+\>$ to Alice, Alice performs a $\sigma _{z}$ rotation on it
and returns it to Bob, who also performs a $\sigma _{z}$ on it.
The total transformation $U$ on the qubit is
\begin{equation}\label{U}
U=\sigma_z\sigma^A_z=\sigma_z R^{\dagger}\sigma_z R=\left(
\begin{array}{cc}
\cos \theta  & -e^{i\phi }\sin \theta  \\
e^{-i\phi }\sin \theta  & \cos \theta
\end{array}%
\right).
\end{equation}
Note that
$
U^{m}=\left(
\begin{array}{cc}
\cos m\theta  & -e^{i\phi }\sin m\theta  \\
e^{-i\phi }\sin m\theta  & \cos m\theta
\end{array}
\right). $ We imagine a procedure wherein a single spin is
exchanged back and forth $2^{j-1}$ times, with Alice and Bob each
applying $\sigma_z$ rotations, such that $U^{2^{j}}$ is performed
on it. A measurement now yields the outcome 1 with probability:
\begin{eqnarray*}
P_{1}&=&\tfrac{1}{2}[1+\cos (2^{j}2\pi T)] =\tfrac{1}{2}[1+\cos (2^{j}2\pi \,0\cdot t_{1}t_{2}t_{3}t_{4}\ldots )] \\
&=&\tfrac{1}{2}[1+\cos ((2\pi \,t_{1}t_{2}\ldots t_{j})+(2\pi
\,0\cdot
t_{j+1}t_{j+2}\ldots ))] \\
&=&\tfrac{1}{2}[1+\cos (2\pi \,0\cdot t_{j+1}t_{j+2}\ldots )]
\end{eqnarray*}
We are therefore back to the situation discussed above for
estimating $t_1$ (although obviously with more exchanges of the
qubit necessitated). As before, we imagine the process is repeated
$n$ times, such that Bob obtains an estimate  $P_{1}^{\prime }$ of
$P_{1}$. The Chernoff bound tells us that the probability the
difference between $P_{1}^{\prime }$ and the true value $P_{1}$ is
greater than some precision $\delta,$ decreases exponentially with
$n.$ That is,
\[
\Pr [\left| P_{1}^{\prime }-P_{1}\right| \geq \delta ]\leq
2e^{-n\delta ^{2}/2}.
\]
By setting $\delta =1/4,$ we obtain a bound that corresponds to
$P_{1}^{\prime }$ agreeing with $P_{1}$ to the first bit - which
in this generalized scenario means that Bob obtains the bit
$t_{j+1}$.  We can therefore bound $n$ as follows:
\[
2e^{-n/32}\leq \epsilon /k\rightarrow n\geq 32\ln (2k/\epsilon ).
\]
The total amount of qubit communication required to obtain bits
$t_1$ through $t_k$ by this procedure is
\[
N= n\times \sum_{j=1}^{k}2^{j-1}=n(2^{k}-1)=O(2^{k}\ln
(2k/\epsilon )).
\]
Note that, since we determine the bits of $T$ independently with
this protocol, the number of \emph{rounds} of communication can be
reduced by running the procedure in parallel. In order to obtain
the other Euler angles accurate to $k$ bits, or, for that matter,
to fix a direction in space with $\theta ,\phi $ angles fixed to
$k$ bits, we can extend this protocol by changing the
transformations that Alice and Bob perform (and/or the initial
state Bob prepares). However this clearly only increases the
communication overhead by a constant factor.

To facilitate comparison with previous work which focussed on
maximizing the average fidelity, we imagine that Alice and Bob use
a variant of the above protocol to obtain, with probability
$(1-\epsilon)^2$, angles $\tilde{\theta},\tilde{\phi}$ which are
``$k$-bit'' estimators of the angles $\theta ,\phi$ specifying
$\vec{n}_A$ (i.e. $|\theta-\theta'|\le 2\pi/2^{k+1},
|\phi-\phi'|\le 2\pi/2^{k+1} $). We have then that $\vec{n}_A\cdot
\vec{n}_{e}\equiv \cos \Delta \alpha\ge
1-\left(\frac{2\pi}{2^k}\right)^2$. (This follows because
$\Delta\alpha\le |\theta-\theta'|+|\phi-\phi'|$ and $\cos x\ge
1-x^2$.) Thus, the choice of $\tilde{\theta},\tilde{\phi}$ leads
to a worst case fidelity of
\[
 F=(1-\epsilon )^2\tfrac{1}{2}(1+\cos
\Delta \alpha )\ge(1-\epsilon )^2(1-2\pi^2\frac{1}{2^{2k}})
\]
(we underestimate the fidelity by assuming that when an error
occurs, then the fidelity of the choice of
$\tilde{\theta},\tilde{\phi}$ is 0 - i.e. worse than random
guessing.) If we take $\epsilon =1/2^{2k},$ then the total qubit
communication is $ N=O(k2^{k})$ (ignoring terms logarithmic in $k$
) while the worst case fidelity is
$F=1-O(\frac{1}{2^{2k}})=1-O(\frac{\log^2 N}{N^{2}})$.

It is useful to understand the above protocol in quantum
computational terms. In effect, Alice and Bob are performing a
combination of a distributed quantum search algorithm \cite{lov}
and a phase estimation algorithm. In a quantum search algorithm, a
generic transformation of the form $(I_t R^\dagger I_{\bar{0}}
R)$, where $R$ is an arbitrary unitary transformation and
$I_t,I_{\bar{0}}$ are phase inversions about source and target
states, is repeated some large number of times in order to
coherently drive the state of the computer. Here we are performing
a similar procedure, where the computer is now only a single bit,
the phase inversions are Alice and Bobs' local $\sigma_z$
rotations, and the unitary transformation $R$ is passively
provided by their lack of a SRF. We may also interpret this
procedure as one in which the eigenvalues of $U$ are being
`quantum computed' - in fact there is much in common here with
Kitaev's version of the quantum phase estimation procedure
\cite{kitaev}.

 We note that a more general
distributed quantum computation would require Alice and Bob to
create entangled states. Without a SRF however, this is at first
glance problematic - since pure entangled states in Alice's frame
are generally mixed in Bob's frame. A possible resolution is for
Alice and Bob to use the encodings of spin states presented in
\cite{BRS}. Such encodings allow for three entangled spin-1/2
particles to form logical qubit states, $\{|0_L\>,|1_L\>\}$ which
are \emph{not} reference frame dependent. As such, Alice and Bob
could, for instance, run the more standard phase estimation
algorithm \cite{NC}, which involves using the discrete fourier
transform to obtain the best $k$ bit estimator of the
eigenvalue(s) of a unitary transformation. Explicitly, the
eigenvalues $e^{\pm i\theta}$ of $U$ (Eq. (\ref{U})), could be
computed as follows: Bob prepares a set of qubits in the state
\[
|\psi\>=\sum_{j=0}^{2^x-1} |j_L\> \otimes |z^+\>.
\]
The subscript $L$ indicates the integers $j$ are encoded in
spin-1/2 systems using the aforementioned binary ``logical''
states about which Alice and Bob both agree despite no SRF; the
number $x$ is a function of $k$ - the number of bits to which we
wish to approximate $\theta$. In the phase estimation algorithm a
series of controlled-$U^{2^j}$ operations are performed on the
second register (the single qubit) controlled on the first
register (the logical qubits). In this communication scenario,
performing these transformations clearly requires the exchange of
the subset of logical qubits being used for the control, as well
as the single spin upon which Alice and Bob perform
controlled-$\sigma_z$ operations. (The state $|z^+\>$ is not an
eigenstate of $U$, as is generally used in the quantum phase
estimation algorithm, however it is an equi-weighted superposition
of the two eigenstates, and this is sufficient - see e.g.
\cite{NC} for details). In the standard manner the phases $e^{\pm
i2^{j} \theta}$ accumulated on the single spin-1/2 are ``kicked
back'' in front of the logical qubit states, and consequently a
discrete fourier transform by Bob on the logical qubit states
will, with probability $(1-\epsilon)$, reveal the best $k$ bit
approximation to $\theta$ providing we choose
$x=k+\lceil\log(2+1/2\epsilon)\rceil$ \cite{NC}. In terms of the
previous discussion regarding direction finding and the fidelity,
this procedure can be shown to give a worst case fidelity that
goes as $F= 1-O(\tfrac{1}{N^2})$, with $N$ the total qubit
communication. Note, however, that this procedure \emph{does}
require the ability to create and exchange large entangled states.

We conclude with some more general observations regarding the
communication complexity of establishing a SRF. We know of no
examples where the ability to exchange classical information
provably helps in reducing the amount of qubit communication
required. It does, as remarked upon in the introduction,
facilitate certain types of protocols in which entanglement might
be ``traded in'' for a reference frame. We leave the reader with
the following related and important question: To what extent does
sharing of one type of reference frame (e.g. synchronised clocks)
facilitate in establishing a different type of reference frame
(e.g. aligned spatial axes). Surprisingly, it seems that in some
cases such facilitation is possible. Consider, for example, the
case when Alice and Bob have synchronised clocks and thus can
quantum communicate perfectly using two (possibly degenerate)
energy eigenstates $\{|e_1\>, |e_2\>\}$ of some system. They can
use a register of these qubits to take the place of the ``logical
qubits'' discussed above in the phase estimation procedure. This
results in a constant factor improvement in the total amount of
qubit communication required.

\begin{acknowledgements} TR would like to thank the Technion for their
hospitality during a visit in which his interest in this problem
was re-ignited. This work is supported by the NSA \& ARO under
contract No. DAAG55-98-C-0040.
\end{acknowledgements}




\end{document}